\documentstyle[12pt]{article}

\begin{document}

\baselineskip = 24pt

\hoffset = -1truecm
\voffset = -2truecm

\title{\bf
Extended  analysis  of  the  "penguin"
part of $K\rightarrow\pi\pi$ amplitude
}

\author{{\bf A.A.Penin}\\
{\it Institute for Nuclear  Research of the Russian Academy of Sciences},\\
{\it Moscow 117312, Russia}\\
{\rm and}\\
{\bf A.A.Pivovarov}\thanks{On leave from Institute for Nuclear
Research, Moscow, Russia}\\
{\it National Laboratory for High Energy Physics (KEK)},\\
{\it Tsukuba, Ibaraki 305, Japan}}
\maketitle

\begin{abstract}
We  make an attempt to clarify the role of the
annihilation or "penguin" mode in the  description
of the $K\rightarrow\pi\pi$ decay within the
Standard Model.   The attention is concentrated on
new operators in the effective $\Delta S=1$
Hamiltonian and the violation of factorization
for mesonic  matrix  elements of the local
four-quark  operators.  We propose a regular
method  to  evaluate the mesonic matrix elements
of $K\rightarrow\pi\pi$ transitions based  on
studying three-point correlators via  {\it  QCD}
sum rules using the chiral effective  theory as
an underlying  low-energy model  for strong
interaction. Matrix elements of the {\it QCD}
penguin operator are calculated within this
approach. The total "penguin" contribution is
found to be relatively large that improves the
theoretical description of the $\Delta I=1/2$ rule
in non-leptonic kaon decays.
\end{abstract}

\vspace{0.5in}
\begin{center}
PACS number(s): 13.25.+m, 11.50.Li, 11.30.Rd, 12.38.Bx.
\end{center}

\thispagestyle{empty}

\newpage

\section{
Introduction.
}

At  present  the   Standard     Model  (SM)
\cite{1} of electroweak  and  strong  interactions
seems   to fit  successfully  all  known  data  of
particle phenomenology  and provides a good
prototype of future  theory of unification.   Even
though the  SM itself  is a  little unsatisfactory
from the aesthetic point of view, it plays an
essential role as a convenient tool of compact
description of all available experimental facts.
However along with many achievements of the SM
there exist several subtle points still to be
understood.  These points are of particular
interest because their resolution might lead to a
discovery of new features absent within the SM
itself and could serve as an indication on new
physics beyond the SM.  Before claiming the
discovery of new physics and demanding going
beyond the SM, however, the corresponding
phenomenon should be examined thoroughly to
guarantee it cannot be explained at the level of
the SM and the proper accuracy is achieved.
Unfortunately this task cannot presently be
easily accomplished.

One example of these subtle points within the SM
is the origin for considerable enhancement of the
$\Delta I=1/2$ parts of the amplitudes in
non-leptonic kaon decays.  Though the $\Delta
I=1/2$ rule is explained qualitatively by  the
influence of  strong interactions the detailed
quantitative description is still lacking.  The
closely related problem of the "direct" {\it CP}
violation in kaon decays  also requires a more
precise quantitative analysis.

Recently numerous attempts have been  made to
improve the theoretical description of
non-leptonic kaon decays in SM with sufficient
accuracy. The efforts are exerted both to extend
the perturbative {\it QCD} analysis \cite{2}
beyond the leading order and to account more
 adequately for long-distance effects using some
models of strong interactions at low energy such
as, for example, chiral  Lagrangians \cite{3},
$1/N_c$ expansion of many color {\it QCD}
\cite{4,5} or lattice simulations \cite{6}.
However the present results still do not reproduce
the experimental data and the existing accuracy
is not satisfactory.

In the present paper we study some questions
connected with the annihilation mode of the
$K\rightarrow\pi\pi$ decay which were not
considered in the previous analysis. This mode is
important in explanation of the $\Delta I=1/2$
rule because it is purely of the $\Delta I=1/2$
sort and its presence is the main distinction
between $\Delta I=1/2$ and $\Delta I=3/2$
amplitudes at least within the perturbation
theory. It also generates the imaginary part of
the effective Hamiltonian and, therefore, is
responsible for the non-zero value of the
$\epsilon'$ parameter describing the direct {\it
CP} violation in kaon decays

At the fundamental level of the SM the strangeness
changing transitions with $\Delta S=1$ occur via
the $W$-boson exchange  between two weak charged
currents.  The short-distance analysis of the
product of weak hadronic currents after removing
the $W$-boson and the heavy quarks results in an
effective $\Delta S=1$ Hamiltonian of the
following  form \cite{7,8,9}
$$
H_{\Delta S=1}= {G_F\over\sqrt
2}V_{ud}V^*_{us}\sum^{6}_{i=1}
[z_i(\mu)+\tau y_i(\mu)Q_i]+h.c.
\eqno(1.1)
$$
Here $G_F$ is the Fermi constant, $V$ stands for
the quark flavor mixing matrix,
$\tau=-{V_{td}V^*_{ts}/V_{ud}V^*_{us}}$,
$z_i(\mu)$ and $y_i(\mu)$ are the coefficients of
the Wilson expansion subtracted at the point $\mu$
and $\{Q_i|i=1,...,6\}$ is the basis of the local
operators containing light quark fields $(u,d,s)$
only with dimension six in mass units (four-quark
operators). In eq.~(1) we have omitted the
contributions of electroweak penguin operators
\cite{9,10}.  The current-current operators
$\{Q_i|i=1,2\}$ do not form a close set under
renormalization and additional so-called {\it QCD}
penguin operators $\{Q_i|i=3,...,6\}$ are produced
by an annihilation diagram.  Since the discovery
of the annihilation mode \cite{7} it was widely
discussed in the literature. The present analysis
however cannot be considered a complete one.
Following points require further investigation:

\noindent
1. The  {\it QCD} perturbation theory  analysis of
the kaon non-leptonic  decay can be improved  by
using more accurate effective Hamiltonian.  In the
standard approach  only the leading terms in the
inverse   masses of heavy quarks  are taken  into
consideration while a proper account of the
non-leading corrections in the inverse mass  of
charmed quark generated by the  penguin-type
diagrams along with the usual expansion in the
strong   coupling constant $\alpha_s$ can be
important.  Analogous corrections to  the
$K^0-\bar K^0$ mixing have  been considered in
ref. \cite{11}  and turn  out not  to be
negligible though  their actual magnitude can
strongly depend on the procedure used for
estimation  of the corresponding matrix element.
Further corrections to the effective Hamiltonian
appear because the top quark is heavier than other
quarks and $W$-boson.  This results in an
incomplete GIM cancellation of the annihilation
diagram including $t$-quark.  Though this question
was widely discussed \cite{9} the appearance of a
new operator in the effective Hamiltonian remained
beyond consideration.

\noindent
2. An achievement of a high precision in
theoretical estimates for the kaon decays stumbles
at a necessity to calculate mesonic matrix
elements of local four-quark operators in the
effective Hamiltonian (1.1). The only method of
computation  entirely based on first principles
seems to be  numerical simulations on the lattice
though so far even this approach has not provided
us  with unambiguous estimates due to some
subtleties connected with the description of
fermions on the lattice.  Meanwhile several
semi-phenomenological techniques have been
developed and applied for computation of those
matrix elements, for example, the many color
expansion of {\it QCD} \cite{4,5}.  Their
precision, however, still need to be essentially
improved.

Recently a new regular method to evaluate the
mesonic matrix elements has been proposed
\cite{p2} where the effective Hamiltonian is
represented in terms of the chiral effective
theory variables and the parameters of the chiral
representation are determined via {\it QCD} sum
rules for an appropriate three-point Green's
function. In
Sect.~3 we briefly describe the method taking as
an example the calculation of the {\it QCD}
penguin operator matrix elements which has been a
subject of a few controversies \cite{4,5,7,19,28}.
This technique will be applied in Sect.~4 to
estimate the scalar gluonium contribution to the
$K\rightarrow\pi\pi$ decay amplitude.

\noindent
3.  The renormalization group improved
perturbation theory does not take into account the
strong interaction of the soft  light quarks and
gluons with the virtual momentum smaller than the
normalization point $\mu\sim 1~GeV$.  The
information on this interaction is entirely
contained in the mesonic matrix elements of the
local four-quark operators.  Well known
 factorization procedure for evaluation of these
matrix elements \cite{12} accounts only for the
"factorizable" part of the interaction \cite{13}.
"Unfactorizable" contributions, for example, those
corresponding to the annihilation of a quark pair
from the four-quark operator into soft gluons are
omitted when the factorization procedure is
performed.  The calculation of these contributions
and the generalization of the matrix elements
beyond the factorization framework can be
systematically done within the approach applied in
Sect.~3 to the calculation of the penguin operator
matrix element.  It turns out to be possible to
obtain an information on the important part of the
factorization violating contribution to the
$K\rightarrow\pi\pi$ decay amplitudes.  This
possibility is connected with the investigation of
a new $K\rightarrow\pi\pi$ decay channel induced
by annihilation of the quark pair from the
four-quark operator into gluons with the
subsequent formation of the pion pair by the soft
gluon cloud, {\it i.e.} the decay channel with the
gluons playing the role of the intermediate state
\cite{rp}.  Being unfactorizable this decay mode
does not appear as a correction to some leading
order contribution and can be studied by its own.
This feature makes obtained results more accurate.

In the Sect.~4 we study a new $K\rightarrow\pi\pi$
decay channel with the simplest scalar colorless
gluon configuration forming an intermediate state.
We calculate both short-distance (perturbative)
and long-distance (non-perturbative) part of the
corresponding amplitude.  The short-distance
analysis is based on the results of the Sect.~2.
To obtain the long-distance contribution the
chiral Lagrangians are used as a low-energy model
of strong interactions and corresponding chiral
coupling is derived via {\it QCD} sum rules.

\section{
The local effective $\Delta S=1$ Hamiltonian
beyond the four-quark approximation.
}

We begin with the calculation of the non-leading
charmed quark mass correction.  Expressions for
Wilson coefficients $z_i(\mu)$ and $y_i(\mu)$ in
eq.~(1.1) in the region of {\it QCD} asymptotic
freedom are obtained by performing the operator
product expansion ({\it OPE}) of two  weak charged
quark currents.  At a typical energy scale of weak
decays of light hadrons these Wilson coefficients
have been with the renormalization group
technique.  After removing all  heavy particles
($W$-boson, $t$-, $b$-, and $c$-quarks) from the
 light sector of the theory formula (1.1)
corresponds to the leading order in inverse masses
of these particles.  The removal of the $c$-quark
however is not very reliable and, in general,
requires a special investigation it is not heavy
enough in comparison with the characteristic mass
scale  in  the sector  of light $u$-, $d$-, and
$s$-quarks, for example, with the $\rho$-meson
mass.  The non-leading terms   in the $1/m_c$
expansion can, therefore, be important  and
require a quantitative consideration.

The effective low-energy tree level Hamiltonian
for $\Delta  S=1$ transitions before decoupling
the $c$-quark reads
$$
H^{tr}_{\Delta S=1}= {G_F\over\sqrt
2}V_{ud}V^*_{us}(Q_2^u-(1-\tau )Q_2^c)+h.c.
\eqno(2.1)
$$
where $Q_2^q=4(\bar s_L\gamma_\mu q_L)(\bar
q_L\gamma_\mu d_L)$, $q_{L(R)}$ stands for
left-(right-) handed quark.  Performing the {\it
OPE} and restricting oneself to the first order
terms  in $\alpha_s$ and $m_c^{-2}$ one can write
the representation for the effective Hamiltonian
in the  form
$$
H_{\Delta S=1}=H^{(6)}+H^{(8)}.
\eqno(2.2)
$$
The  first addendum on the right-hand side ({\it
rhs}) of eq.~(2.2), $H^{(6)}$, corresponds to the
leading contributions in $1/m_c$ and coincides
with the {\it rhs} of eq.~(1.1).  Second addendum
on the {\it rhs} of eq.~(2.2), $H^{(8)}$, is the
leading order $1/m_c$ correction \cite{p1}
$$
H^{(8)}= {G_F\over\sqrt 2}V_{u
d}V^*_{us}(1-\tau){\alpha_s\over 4\pi}
\left(\sum^7_{i=1} C_i^{(8)} Q_i^{(8)}+
\sum^4_{i=1} C_i^{(7)} m_sQ_i^{(7)}\right) +h.c.
\eqno(2.3)
$$
where a basis $\{ Q^{(8)}_i|i=1,...,7\}~ (\{
Q^{(7)}_i|i=1,...,4\})$ of  the  local  operators
with dimension eight (seven) in mass units is
chosen in  the form
$$
Q_1^{(8)} =\bar s_L(\hat DG_{\mu\alpha}G_{\nu\mu}
\sigma_{\alpha\nu}+G_{\nu\mu}\sigma_{\alpha\nu}
\hat DG_{\mu\alpha})d_L,
$$
$$
Q_2^{(8)} =ig_s\bar s_L(J_\mu\gamma_\alpha
G_{\alpha\mu}- \gamma_\alpha
G_{\alpha\mu}J_\mu)d_L,
$$
$$
Q_3^{(8)}
=\bar s_L(P_{\alpha}G_{\mu\alpha}\gamma_\nu
G_{\nu\mu}+\gamma_\nu G_{\nu\mu}
G_{\mu\alpha}P_{\alpha})d_L,
$$
$$
Q_4^{(8)} =g_s\bar
s_L(G_{\mu\nu}\sigma_{\mu\nu}\hat J+ \hat J
G_{\mu\nu}\sigma_{\mu\nu})d_L,
$$
$$
Q_5^{(8)}
=i\bar s_L(G_{\mu\nu}\sigma_{\mu\nu}\gamma_ \alpha
G_{\alpha\beta}P_\beta -P_\beta\gamma_\alpha
G_{\alpha\beta} G_{\mu\nu}\sigma_{\mu\nu})d_L,
$$
$$
Q_6^{(8)} =\bar s_L(D^2\hat J)d_L, ~~Q_7^{(8)}
=i\bar s_L(\hat DG_{\nu\mu}G_{\nu\mu}-
G_{\nu\mu}\hat DG_{\nu\mu})d_L,
$$
$$
Q_1^{(7)}
=\bar s_R(G_{\mu\nu}\sigma_{\mu\nu}
G_{\alpha\beta} \sigma_{\alpha\beta})d_L,~~
Q_2^{(7)}=\bar s_R(G_{\mu\nu}G_{\nu\mu})d_L,
$$
$$
Q_3^{(7)} =i\bar s_R(G_{\nu\alpha}G_{\alpha\mu}
\sigma_{\nu\mu})d_L,~~ Q_4^{(7)}=\bar s_R(J_\mu
P_\mu +P_\mu J_\mu)d_L.
\eqno(2.4)
$$
Here $P_\mu  =i\partial_\mu+g_sA_\mu$ is the
momentum  operator  in  the  presence  of the
external  gluon  field $A_\mu \equiv A_\mu^a t^a$,
$t^a$ are the standard generators of the color
group $SU(3)$, $G_{\mu\nu}\equiv G^a_{\mu\nu}t^a$
is  the gluon field strength tensor, $J_\mu\equiv
\sum_{q=u,d,s} (\bar  q\gamma_\mu t^a q)t^a$ and
$\sigma_{\mu\nu}=i[\gamma_\mu,\gamma_\nu]/2$.
When eq.~(2.3) is derived we consider $u$- and
$d$-quark to be massless, keep the first order
quantities in strange quark mass and use the
equations of motion
$$
\bar  s\hat  P=m_s\bar s,~~~\hat P d=0,
$$
$$
[P_\mu, G_{\mu\nu}]=iD_\mu
G_{\mu\nu}=-ig_sJ_\nu .
\eqno(2.5)
$$
Straightforward calculations give the
following numerical values for the coefficients
$C_i^{(j)}$ to the leading order in the strong
coupling constant $\alpha_s$
$$
C_1^{(8)}={8\over 15},~~C_2^{(8)}=-{16\over 15},
{}~~C_3^{(8)}=-{4\over 5}, ~~C_4^{(8)}={2\over 15},
$$
$$
C_5^{(8)}=0, ~~C_6^{(8)}=-{8\over 15},
{}~~C_7^{(8)}=-{2\over 15},
$$
$$
C_1^{(7)}=-{2\over 5},~~C_2^{(7)}=-{2\over 5},
{}~~C_3^{(7)}={6\over 5},~~C_4^{(7)}=0.
\eqno(2.6)
$$
Thus we have generalized the effective Hamiltonian
for $\Delta S=1$  decays beyond the leading order
in  $1/m_c$.

To complete our treatment of the local effective
Hamiltonian the annihilation of the top quark
should be also considered.  Because the top quark
is heavy enough with respect to the $W$-boson
\cite{14} these two particles should be integrated
out simultaneously to obtain the  local effective
Hamiltonian. The  procedure of decoupling the
heavy top quark is described in details in ref.
\cite{9} but one point has remained beyond the
analysis.  It is connected with other penguin-type
operator $m_sQ^{(5)}=m_s{\bar
s}_Rg_sG_{\mu\nu}\sigma^{\mu\nu}d_L$.  This
quark-gluon operator was omitted in the early
papers \cite{7,8} because the corresponding
coefficient function vanishes in the first order
in $\alpha_s$. The next-to-leading calculations
\cite{15} has verified the smallness of this
Wilson coefficient.  However  the above statement
holds only under the assumption $m_t\ll M_W$, {\it
i.e.} in the leading order in $m_q^2/M_W^2$ when
the complete GIM cancellation takes place. In the
case of a heavy top quark it becomes invalid and
the quark-gluon operator $m_sQ^{(5)}$ appears in
the effective Hamiltonian already in the first
order in $\alpha_s$. This additional contribution
has been obtained in ref. \cite{16} and reads
$$
\Delta H^{(6)}={G_F\over\sqrt
2}V_{ud}V^*_{us}\tau
C^{(5)}(\mu )m_sQ^{(5)}(\mu ),
$$
$$
C^{(5)}(\mu )={1\over 16\pi^2}
\left(F(x_c)-F(x_t)\right)\eta (\mu )
,~~~x_q={m_q^2\over M_W^2},
$$
$$
F({x_q})={1\over 3}{1\over (x_q-1)^4}
\left({5\over 2}{x_q}^4-7{x_q}^3+
{39\over 2}{x_q}^2-19{x_q}+4-
9{x_q}^2\ln{x_q}\right),
$$
$$
F(x_c)\sim F(0)={4\over 3}.
\eqno (2.7)
$$
The renormalization group factor
$$
\eta(\mu )
=\left({{{\bar{\alpha}}_s}(m_b)
\over{{\bar{\alpha}}_s}(M_W)}\right)
^{\gamma^{(5)} /{2\beta_5}}
\left({{{\bar{\alpha}}_s}(m_c)
\over{{\bar{\alpha}}_s}(m_b)}\right)
^{\gamma^{(5)} /{2\beta_4}}
\left({{{\bar{\alpha}}_s}(\mu
)\over{{\bar{\alpha}}_s}(m_c)}\right)
^{\gamma^{(5)} /{2\beta_3}}
$$
where $\gamma^{(5)}=-28/3$ is the anomalous
dimension of the operator $m_sQ^{(5)}$ \cite{17},
$\beta_{n_f}=11-{2\over3}n_f$, $n_f$ is the number
of active quarks flavors, can be easily obtained
because the operator $m_sQ^{(5)}$ does not mix
with the four-quark operators in the leading
order.

The contributions of the $u$- and $c$-quarks to
the real part of the effective Hamiltonian cancel
one another via GIM mechanism and the operator
$m_sQ^{(5)}$ mainly contributes to the imaginary
part of the effective Hamiltonian and, therefore,
can be important in the analysis of direct {\it
CP} violation.  However, its contribution is
suppressed numerically because $\eta(\mu )<1$ and
the function $F(x)$ changes slowly.  Indeed, at
the point $\Lambda_{QCD}=0.3~GeV$, $\mu=1~GeV$,
$m_t=130~GeV$, we have $C^{(5)}=0.0009$, while the
numerical value of the same Wilson coefficient of
the dominant penguin operator
$Q_6=-8\sum_{q=u,d,s}(\bar s_L q_R)(\bar q_R d_L)$
is $y_6=0.102$ \cite{9}.

Thus, the complete form of the effective
Hamiltonian up to the first order in $m_s$,
$1/m_c$ and $\alpha_s$ with $m_t\sim M_W$ becomes
available now (eqs.~(2.2, 2.3, 2.7)). The
electroweak penguin operators have to be also
included into the complete expression.

As an example of using the above Hamiltonian we
consider the $K\rightarrow\pi\pi$ decays.  For
this end we have  to  extract  an  information
about the matrix elements  of  the  local
operators $Q^{(j)}$ between the mesonic states.
We start with the operators $m_sQ^{(7)}_i$ and
$m_sQ^{(5)}$.  These operators contain explicitly
the strange quark mass and, therefore, in the
leading order of the chiral expansion they
correspond to the tadpole term in the chiral weak
Lagrangian.  In detail the tadpoles will be
considered in Sect.~3. Now we only note that
tadpoles  do not generate any observable effect
and can be neglected in the leading order  of the
chiral symmetry  breaking.

Here a remark about the role of the operator
$m_sQ^{(5)}$ in determination of the parameter
$\epsilon'$ is necessary. Recently an enhancement
of the corresponding Wilson coefficient in the
next-to-leading order in $\alpha_s$ has been discovered
and a significance of this operator in the
analysis of the direct {\it CP} violation was
announced \cite{q1}.  However in ref.  \cite{q1}
as in an earlier paper \cite{16} the tadpole
character of the operator $Q^{(5)}$ has not been
recognized and its contribution to the parameter
$\epsilon'$ was strongly overestimated. The
correct treatment of this problem has appeared in
the most recent paper \cite{q2} which is in
agreement with our analysis.

Thus the problem is reduced to the estimation of
the matrix elements  of the operators $Q^{(8)}_i$.
At present there is no regular method to calculate
that kind of object within {\it QCD} excepting the
numerical simulations on the lattice. To estimate
at least the scale of the non-leading $1/m_c$
corrections we will work with the simplified
model.  As a first approximation we take the
operators that survive  after the factorization
procedure.  Then one selects the operators
containing scalar quark  currents which  can be
written as
$$
(\bar s_L
G_{\mu\nu}\sigma_{\mu\nu}q_R)(\bar q_R d_L)
{}~~{\rm and}~~
(\bar s_L q_R)(\bar q_R
G_{\mu\nu}\sigma_{\mu\nu}d_L).
$$

This step  seems to  be justified because in  the
case  of  dimension  six operators  the   similar
"penguin-like" structures are strongly enhanced
and dominate the others (see Sect.~3).  The   last
simplification   consists in the substitution
$$
\bar q g_s G_{\mu\nu}\sigma_{\mu\nu}q
\rightarrow m_0^2\bar q q
\eqno(2.8)
$$
where $m_0$ determines the scale of non-locality
of the quark condensate and is defined by the
equation $\langle \bar q g_s G_{\mu\nu}\sigma_
{\mu\nu}q\rangle =m_0^2\langle \bar qq\rangle ,
{}~~~m_0^2(1~GeV)=0.8\pm 0.2~GeV^2$ \cite{18}.

This
substitution is valid in the chiral limit for the
operator $\bar q g_s G_{\mu\nu}\sigma_{\mu\nu}q$
with dimension five in mass units. We suppose that
it  is justified also in our case at least for
estimates up to the order of magnitude.  Actually
all above assumptions as the factorization
procedure in the case of dimension six operators
become exact within the many color limit of {\it
QCD}, $N_c\rightarrow\infty$, to the leading order
in $N_c$.

Having adopted the  assumptions described above
the only operator which has the non-zero matrix
element for the considered process is the operator
$Q^{(8)}_4$ and the corresponding quantity reads
$$
\langle \pi\pi|Q_4^{(8)}|K\rangle ={m^2_0\over 4}
\langle \pi\pi|Q_6|K\rangle .
\eqno(2.9)
$$

We should note that the coefficients  $C_i^{(8)}$
are finite to the leading order of the $\alpha_s$
and independent of renormalization scheme. We can
therefore use the leading order values of the
mesonic matrix elements that is consistent up to
the used level of accuracy. In the contrast, the
next-to-leading $\alpha_s$ corrections to Wilson
coefficients depends on the renormalization scheme
and to make the physical amplitudes scheme
independent, matching between Wilson coefficients
and mesonic matrix elements in the same
renormalization scheme has to be made \cite{2}.

Thus, taking into  account the first  order
$1/m_c$ corrections is reduced to the effective
shift of  the  coefficients  in front  of  the
penguin operator $Q_6$ in the effective
Hamiltonian (1.1)
$$
z_6\rightarrow
\left(z_6+{\alpha_s\over 4\pi}{m^2_0 \over
4m^2_c}C^{(8)}_4\right),
{}~~~~~y_6\rightarrow
\left(y_6-{\alpha_s\over 4\pi}{m^2_0 \over
4m^2_c}C^{(8)}_4\right).
\eqno(2.10)
$$
Using   the    numerical   values    $z_6=-0.015$,
$y_6=-0.102$ at the point
$\Lambda_{QCD}=0.3~GeV$, $\mu=1~GeV$,
$m_t=130~GeV$  \cite{9}  one  finds numerically
the relative corrections to the Wilson
coefficients in the form
$$
z_6\rightarrow z_6(1-0.1),
{}~~~~~y_6\rightarrow y_6(1+0.01).
$$
The main correction appears in the real part  of
the  Wilson  coefficient  of the penguin operator
$Q_6$.  Parametrically, the contribution of the
dimension eight operators can be as large as one
half ($m^2_0/m^2_c \sim 0.5$) of the one of
dimension six operators, and not too small.  In
fact, we have found that when one estimates the
mesonic  matrix element  of the local operator
with   dimension   eight within the  simplest
factorization framework the non-leading $1/m_c$
contributions  to the kaon decay amplitudes are
about $10\%$ of the leading ones.  However a large
violation of the factorization for the matrix
elements of the dimension eight operators does not
seem to be impossible and the real value of the
non-leading $1/m_c$ correction can be estimated
only when somewhat self-consistent method to
calculate these matrix elements within {\it QCD}
will be available.

\section{
The {\it QCD} penguin operator contribution to
the $K\rightarrow\pi\pi$ decay amplitude.
}

In this section we demonstrate a functioning of
the method developed earlier in ref. \cite{p2} for calculating
a contribution of the penguin operator $Q_6$ to
the $K\rightarrow\pi\pi$ amplitude.  This operator
draws the special interest because its mesonic
matrix element prevails over the
current-current and other {\it QCD} penguin
operators and gives a dominant
contribution to the parameter $\epsilon'$ (we do
not consider the electroweak penguins).  Moreover,
the estimates of matrix elements of the
current-current operators are much less
controversial and are quite reliably given by the
simple factorization procedure. We leave aside
technical details of the calculations that can be
found in ref.  \cite{p2} and focus on the
main features of the approach.

The matrix element of the $K\rightarrow\pi\pi$
decay amplitude due to the operator $Q_6$ only
reads
$$
\langle \pi^+\pi^-|Q_6(\mu)|K^0\rangle =\langle
\pi^0\pi^0|Q_6(\mu)|K^0\rangle
\equiv \langle \pi\pi|Q_6(\mu)|K^0\rangle .
\eqno(3.1)
$$
For comparison with other approaches we  use the
following parametrization of this matrix element
$$
{1\over i}
\langle \pi\pi|Q_6(\mu)|K^0\rangle =-B_pf_Km^2_K,
{}~~~~f_K=1.23f_\pi,
{}~f_\pi=132~MeV~\cite{14}
\eqno(3.2)
$$
where the $f_Km^2_K$ factor fixes a natural mass
scale while $B_p$ is a dimensionless parameter to
be computed.

We start with constructing the chiral effective
theory describing weak  and strong
interactions of pseudoscalar mesons at
low energy \cite{3,20}. The operator $Q_6$ is a composite
quark operator which can be define accurately in
the asymptotic freedom regime
within perturbation theory using the proper
prescription of renormalization.
Its matrix element can be hardly found in the
practically interesting case of low energies
where confinement takes place. However one can
try to define the operator $Q_6$ at low
energies within chiral perturbation theory,
{\it i.e.} to find an effective realization
of the operator $Q_6$ in terms of mesonic
variables. The operator $Q_6$ belongs to the
$8_L\times 1_R$ irreducible representation of
chiral $SU(3)_L\times SU(3)_R$ group and has the
following $SU(3)_V$ flavor quantum numbers:
$S=1,~I=1/2,~I_3=-1/2$.  To the lowest order of a
chiral perturbation theory expansion there are two
Lorentz invariant bosonic operators with relevant
properties $(\partial_\mu U^\dagger \partial^\mu
U)_{23}$ and $(\chi^\dagger U+U^\dagger
\chi)_{23}$ where $\chi$ stands for the
pseudoscalar meson mass matrix and
$U=e^{-i{\sqrt{2}\phi/f_\pi}}$ is the unitary
matrix describing the octet of pseudoscalar
mesons.  Hence the operator $Q_6$ can be
represented as
$$
Q_6=-f_\pi^4[g(\partial_\mu
U^\dagger  \partial^\mu U)_{23}+
g'(\chi^\dagger U+U^\dagger \chi)_{23}]
\eqno(3.3)
$$
where $g$ and $g'$ are the dimensionless
parameters which are not fixed by symmetry
requirements.  After such an {\it ansatz} the
mesonic matrix elements of the operator $Q_6$
become exactly computable within the effective
theory, in other words, we have found a suitable
kinematical framework for our dynamical problem
which consists in determination of $g$ and $g'$.
For our aim we need the  matrix elements
$$
\langle \pi\pi|Q_6|K^0\rangle =4igf_\pi m^2_K,
\eqno(3.4a)
$$
$$
\langle
\pi^+(p_1)|Q_6|K^+(p_2)\rangle = f^2_\pi
(4g(p_1p_2)+4g'm^2_K),
\eqno(3.4b)
$$
$$
\langle
0|Q_6|K^0\rangle =-4ig'f^3_\pi m^2_K.
\eqno(3.4c)
$$
We put the matrix element in eq.~(3.4a) on the
meson mass shell  because this corresponds to
the physical amplitude of  $K\rightarrow\pi\pi$
transition. We also take into account that the term
proportional to $\chi$ is equal to a full
derivative due to equations of motion and does not
contribute to the amplitude at zero momentum
transfer.  The matrix elements in eqs.~(3.4b) and
(3.4c) are some auxiliary amplitudes considered
only for determination of $g$ and $g'$ and can be
computed at arbitrary point since they are
momentum independent to the leading order of the
chiral expansion.  So, we do not include the
contribution of the so-called tadpole term
$(\chi^\dagger U+U^\dagger \chi)_{23}$ to the
physical amplitude (3.4a).  As is well known the
appearance of such a term is a consequence of
working with a wrong vacuum solution. This term
merely renormalizes the strong effective
Lagrangian (beyond the mass shell) and can be
absorbed into the meson mass matrix by a suitable
$SU(3)_L\times SU(3)_R$ rotation. Thus, the
tadpoles do not generate any observable effects
for the effective chiral Lagrangian of order $p^2$
to the first order in $G_F$ \cite{3,21}.

The problem is reduced to the determination of the
single parameter $g$. It can be done by studying
an appropriate three-point Green's function (GF)
via the sum rules technique. We choose the GF in
the form
$$
G_\mu (p,q)=i^2\int \langle 0|Tj^5_\mu (x) Q_6(0)
j^5(y)|0\rangle e^{ip_2x-ip_1y}dxdy=
$$
$$
=i^2\int \langle 0|Tj^5_\mu (x) Q_6(y) j^5
(0)|0\rangle e^{i(p-{q\over 2})x+iqy}dxdy
\eqno(3.5)
$$
where $p_1=p+q/2$, $p_2=p-q/2$ and we take
$j^5_\mu=\bar d\gamma_\mu\gamma_5 u,~j^5=\bar
u\gamma_5 s$ as  interpolating operators for pion
and kaon fields.  Because of
dispersion relations, GF (3.5) looks like
$$
{\langle j^5_\mu|\pi^+(p_2)\rangle \langle \pi^+(p_2)
|Q_6|K^+(p_1)\rangle \langle K^+(p_1)|j^5\rangle
\over p_2^2(p_1^2-m^2_K)}+{R_L\over p_2^2}+
{R_R\over p_1^2-m_K^2}+\ldots
\eqno(3.6)
$$
where the ellipsis stands for states without any
kaon or pion poles, the quark currents projections
on the mesonic states are
$$
\langle 0|j^5_\mu|\pi^+(p)\rangle =if_\pi p_\mu,~~~
\langle K^+(p)|j^5|0\rangle =-i{f_K m_K^2\over m_s}
\eqno(3.7)
$$
and we work with the massless pion ($u$-,
$d$-quarks).  The representation (3.6) needs a
comment.  The matter is that to extract
information about the matrix element in the first
addendum in eq.~(3.6) one has to distinguish the
resonance contribution from miscellaneous ones.
It is possible because, contrary to all other
states, the resonance leads to the double pole in
the dispersion relation.  In representation (3.6)
it can be achieved within the kinematics where
$(pq)=m^2_K/2$. Setting $(pq)=m^2_K/2$ and
multiplying eq.~(3.6) by the expression
$(p^2+q^2/4-m^2_K/2)$ one obtains
$$
{\langle j^5_\mu|\pi^+(p_2)\rangle \langle \pi^+(p_2)
|Q_6|K^+(p_1)\rangle \langle K^+(p_1)|j^5\rangle
\over (p^2+{q^2\over 2}-{m^2_K\over 2})}+R_L+R_R
+\ldots
\eqno(3.8)
$$
where the resonance contribution is explicitly
distinguished as  a pole term. Eq.~(3.8) can now
be treated by the standard sum rules technique.

Straightforward calculations give the following
asymptotic expansion for the GF
$$
G_\mu (p,q)=p_{2\mu}G(p,q)+\ldots
$$
and
$$
G(p,q)=
-{3\over 2\pi^2}{2(pq)\over p^2}\ln\left({-p^2\over
\mu^2}\right) \langle \bar\psi\psi\rangle +
{3\over 4\pi^2}\gamma\ln\left({-p^2\over
\mu^2}\right)\langle \bar\psi\psi\rangle +
$$
$$
+O(p^{-6})+({\rm bilocal~part})
\eqno(3.9)
$$
where $\langle \bar uu\rangle =\langle \bar
dd\rangle \equiv \langle \bar\psi\psi\rangle $.
We  work in the leading approximation of the
$q^2$ power expansion and simply put $q^2=0$.  At
the same time we have to keep the first order
terms in the scalar product $(pq)=m^2_K/2$ and the
strange quark mass $m_s$ and we also have to
distinguish the strange quark vacuum condensate
from the non-strange one $\gamma ={\langle \bar
ss\rangle  / \langle \bar\psi\psi\rangle }-1\neq
0$ since this difference represents a parameter of
$SU(3)_V$ symmetry breaking.  As we see all terms
of the zeroth order in $m_s$ cancel themselves in
eq.~(3.9) in agreement with the chiral structure
of the operator $Q_6$.  The bilocal part of {\it
OPE} does not contribute to the parameter $g$ to
the $O(p^{-6})$ order due to a specific
non-symmetrical choice of the
the interpolating currents in eq.~(3.5) \cite{p2}.

To determine the parameter $g$ we apply the finite
energy sum rules technique \cite{24} to the
function $G(p^2)(p^2-{m^2_K/2})$ which is defined
by the equation
$$
G(p^2)=G(p,q)\mid_{q^2=0,~(pq)={m^2_K\over 2}}.
\eqno(3.10)
$$
The result for the matrix element $\langle
\pi^+|Q_6|K^+\rangle $ reads
$$
\langle \pi^+|Q_6(s_0)|K^+\rangle =
{3\over 2\pi^2}{m_s\langle \bar\psi\psi\rangle \over
f_Kf_\pi}s_0\left(1+{s_0\gamma\over 4m^2_K}\right)
\eqno(3.11)
$$
where $s_0$ is a duality interval. Combining
eq.~(3.11) with the representation (3.4b) and
using the {\it PCAC} relation $-2m_s\langle
\bar\psi\psi\rangle =f^2_Km^2_K$ we find
$$
g+2g'=-{3\over 8\pi^2}{f_K\over f\pi}{s_0\over
f^2_\pi}\left(1+{s_0\gamma\over 4m^2_K}\right).
\eqno(3.12)
$$
It can be easy shown \cite{p2} that the first
addendum in brackets on the {\it rhs} of
eq.~(3.12) represents exactly the contribution of
the constant $g$ to the entire sum. Thus we have
$$
g=-{3\over 8\pi^2}{f_K\over f\pi}{s_0\over f^2_
\pi}.
\eqno(3.13)
$$
The question now is which value for the duality
interval $s_0$ has to be used.  Throughout the
calculation we keep only the first order quantities
with respect to  $SU(3)_V$ symmetry breaking
parameter -- the strange quark
mass; in eq.~(3.11) the chiral suppression
is present as a factor $m_s$.  Hence, for
consistency, we should use for the duality
interval $s_0$ its  value in the chiral limit,
{\it i.e.} the value of the pion duality interval
$s^\pi_0=0.8~GeV^2$ \cite{25}.  There exists a
prejudice that the duality interval in the
pseudoscalar channel can be abnormally large due
to a possible contribution of so-called "direct"
instantons \cite{26} but the real estimate of that
contribution is practically absent while the use
of the above value reproduces the pion decay
constant $f_\pi$ with reasonable accuracy. Actual
value of the $s_0$ parameter can be found only
after adding the non-perturbative corrections due
to higher dimension operators which are supposed
to be small in our case.  Finally for the
parameter $B_p$ we obtain
$$
B_p(s_0^\pi ) ={3\over
2\pi^2}{s_0^\pi\over f^2_\pi}=7.
\eqno(3.14)
$$
Eq.~(3.14) is written for the normalization point
$\mu^2=s_0^\pi$.  For an arbitrary normalization
point we have
$$
B_p(\mu^2)=B_p(s_0^\pi )\left({\alpha
(s_0^\pi
)\over\alpha(\mu^2)}\right)^{-{\gamma_6/2\beta_3}}
\eqno(3.15)
$$
where $\gamma_6=-14$
is the anomalous dimension of the operator $Q_6$.

The strong dependence of the {\it rhs} of
eq.~(3.14) on the parameter $s_0$ is considerably
smoothed by the renormalization group factor
$({\alpha (s_0)\over\alpha (\mu^2)})^{7/9}$.
Indeed, the result for the parameter $B_p(1~GeV)$
changes only from $8.19$ to $9.38$, {\it i.e.}
less than $15\%$, when one substitutes the kaon
duality interval $s_0^K=1.2~GeV^2$ \cite{25} into
eqs.~(3.14, 3.15) (the {\it QCD} scale is chosen
to be $\Lambda_{QCD}=0.3~GeV$).

Let us now estimate the uncertainty of our result.
On the physical side the errors originate from the
higher order terms in chiral expansion which are
omitted in eq.~(3.3). Their relative weight can be
represented by the ratio
$m_K^2/\Lambda^2_\chi ,$
where $\Lambda_\chi$ is a chiral scale parameter.
Both on theoretical and empirical grounds one
expects \cite{27}
$$
\Lambda^2_\chi =8\pi^2f_\pi^2\sim 1~GeV^2.
\eqno(3.16)
$$
Thus, a choice of the representation for a
quark-gluon operator (for example, $Q_6$) to the
leading order in the chiral effective theory might
introduce a 25\% error in the estimate of the real
physical matrix element.

On the theoretical side of the sum rules the
errors come from several sources:

\noindent
1. Corrections due to operators with higher
dimensionality which start with the term $\langle
\bar\psi{g_s^2}G^2\psi\rangle {(pq)/p^6}$ and,
probably, some  $O(p^{-6})$ bilocal contribution.
They are under control and do not lead to any
sizable variation of our results though their
actual magnitude is questionable because it
requires making some estimates of the vacuum
expectation values of higher dimension and bilocal
operator;

\noindent
2. Perturbation theory corrections to the
coefficient functions of the leading operators.
The correct inclusion of these correction requires
a complete generalization of the effective
Hamiltonian up to the $\alpha_s^2$ order.
Therefore we consequently work only with the first
order in $\alpha_s$ terms in the effective
Hamiltonian and omit radiative correction to the
non-leading operator $Q_6$ since its coefficient
function is already of the $\alpha_s$ order.

Thus, we have calculated the  matrix element of
the penguin operator via the {\it QCD} sum rules
with the result
$$
{1\over i}
\langle \pi\pi|Q_6(1~GeV)|K^0\rangle =
-(0.34\pm 0.09)~GeV^3
\eqno(3.17)
$$
or, in terms of  the parameter $B_p$
$$
B_p(1~GeV)=8.2\pm 2.1
\eqno(3.18a)
$$
where the error bars estimate contributions of
higher orders of the chiral expansion.  The
analysis carried out above shows that the
uncertainties coming from other sources are
relatively small. This result can be important for
the precise calculation of the $\epsilon'$
parameter.

The value $(3.18a)$ is somewhat larger (with a
factor $1.3$ when $m_s(1~GeV)=175~MeV)$ than
the leading order result obtained within the
$1/N_c$ expansion framework \cite{5}
$$
B_p(1~GeV)=6.4\left({175~MeV\over
m_s(1~GeV)}\right)^2.
\eqno(3.18b)
$$
However a sizable enhancement (factor $1.5-2.0$)
of the leading order result (eq.~(3.18b)) due
to the next-to-leading corrections in the $1/N_c$
expansion has been found \cite{28}.
This observation is in good agreement with our
calculations. Our result gets also into the
interval given by lattice models \cite{19}
$$
B_p(1~GeV)=11\pm 3
\eqno(3.18c)
$$
but it is
somewhat smaller than optimistic estimate of
ref. \cite{7} where the soft pion technique was
used
$$
B_p(1~GeV)=
12\left({175~MeV\over m_s(1~GeV)}\right)^2.
\eqno(3.18d)
$$

It has been also shown that the Wilson coefficient of
the operator $Q_6$ is essentially increased by the
next-to-leading $\alpha_s$ terms \cite{2}. For
example,  at the point  $\Lambda_{QCD}=0.3~GeV$,
$\mu=0.8~GeV$, the numerical value of the leading
order coefficient $z^{(l.o.)}_6=-0.028$ while
inclusion of the next-to-leading $\alpha_s$
corrections leads to the result
$z^{(n.l.)}_6=-0.098$. However the consistent
generalization of the effective Hamiltonian beyond
the leading logarithmic approximation requires to
derive the corresponding mesonic matrix elements
up to the same order in $\alpha_s$ (see Sect.~2).
Thus this enhancement in general can be canceled
by radiative corrections to the matrix element of
the operator $Q_6$. If it does not occur the penguin
operator becomes quite important in the analysis
of $\Delta I=1/2$ rule since it provides
approximately $20\%$ of the physical value of the
decay amplitude with the isotopic spin transfer
$\Delta I=1/2$.

\section{
The scalar gluonium contribution to
the $K\rightarrow\pi\pi$ decay amplitude.
}

As it has been already  pointed out the possible
source of the enhancement of the $\Delta I=1/2$
amplitude is the unfactorizable contributions to
the matrix elements of the local four-quark
operators produced by the low-energy strong
interaction of the light quarks. In the chiral
Lagrangians approach these non-factorizable
contributions reveal themselves in two different
ways:  first, they appear as corrections to the
couplings characterizing the "factorizable" weak
chiral Lagrangian in $O(p^2)$ and higher orders,
second, some new non-factorizable terms emerge.
The latter is the case for $K\rightarrow\pi\pi$
decay mode with gluons forming an intermediate
state.

The unfactorizable corrections of the first and of
the second type have already been discussed in the
framework of $1/N_c$ expansion \cite{4,5}. In ref.
\cite{4} the non-perturbative corrections to the
weak chiral Lagrangian of the leading $O(p^2)$
order caused by the presence of the
non-perturbative gluon condensate was derived
while in ref. \cite{5} the $O(p^4)$ order
contribution of the unfactorizable chiral loops
was computed and the sizable violation of the
factorization has been found. However in both
pointed approaches the above mode was omitted.
Meanwhile it is quite reasonable to suppose that
the gluons being considered as exchange particles
can be important in explanation of the $\Delta
I=1/2$ rule because they carry a zero isospin and
contribute only to the $\Delta I=1/2$ amplitude.
Moreover that kind of contribution being
non-leading in chiral expansion nevertheless can
be sizable due to the strong effects of the
non-perturbative gluon vacuum.

Before treating the long-distance effects of the
meson-gluon transitions it is very useful to
consider the similar phenomenon arising already in
perturbative {\it QCD} as a leading correction in
the inverse mass of the charmed quark given by
eq.~(2.3). If we restrict the analysis to the
scalar colorless gluon configuration
$G^a_{\mu\nu}G^a_{\mu\nu}$ only the operators
$Q^{(8)}_3,~m_sQ^{(7)}_1$, and $m_sQ^{(7)}_2$ give
a contribution and eq.~(2.3) gets the  form
$$
H^{(8)}_G={G_F\over\sqrt 2}V_{ud}V^*_{us}(1-\tau)
\left(-{1\over 120}{1\over m_c^2}m_s\bar s_Rd_L
{\alpha_s\over\pi}G^a_{\mu\nu}G^a_{\mu\nu}\right)
+h.c.
\eqno(4.1)
$$
To derive the effective chiral Lagrangian
involving Goldstone degrees of freedom only which
then can be used for the calculation of the decay
amplitude one has to replace the  {\it QCD}
operator
$$
m_s\bar s_Rd_L{\alpha_s\over\pi}
G^a_{\mu\nu}G^a_{\mu\nu}
\eqno(4.2)
$$
in eq.~(4.1) by its mesonic realization. Using its
chiral transformation properties one can write
down the representation
$$
m_s\bar s_Rd_L{\alpha_s\over\pi}
G^a_{\mu\nu}G^a_{\mu\nu}=Af_\pi^6(U^\dagger\chi
)_{23}+
Bf^4_\pi (U^\dagger\chi
)_{23}tr_{fl}(\partial_\mu U^\dagger
\partial^\mu U)+
$$
$$
+({\rm other}~O(p^4)~{\rm terms})+O(p^6)
\eqno(4.3)
$$
where  $A$ and $B$ are the dimensionless
parameters.  The second term on the {\it rhs} of
eq.~(4.3) is separated from the other $O(p^4)$
structures for the reason will be clarified below.
The $O(p^2)$ term in  eq.~(4.3)  is exactly a
tadpole term which should be omitted.  So the
contribution to the physical amplitude is
determined by the $O(p^4)$ part of the chiral
representation (4.3). The most transparent way to
obtain this part is to consider the quark-gluon
operator (4.2) as a product of the (pseudo)scalar
quark current and scalar colorless gluon operator.
Then one can replace the quark current by its
mesonic realization according to the {\it PCAC}
hypothesis and current algebra \cite{20}
$$
m_s\bar s_Rd_L\rightarrow
-{f^2_\pi\over 8} (U^\dagger\chi )_{23}.
\eqno(4.4)
$$
On the other hand there is a low-energy theorem
based on the fundamental properties of the
energy-momentum tensor which gives the chiral
representation of the gluon operator \cite{22,29}
$$
{\alpha_s\over\pi}G^a_{\mu\nu}G^a_{\mu\nu}=
-{2\over\beta_3}f^2_\pi tr_{fl}(\partial_\mu U^\dagger
\partial^\mu U)+O(p^4).
\eqno(4.5)
$$
Eqs.~(4.4, 4.5) allow us to determine the parameter
$B$
$$
B={1\over 4\beta_3}
\eqno(4.6)
$$
which now modulates the unique term of $O(p^4)$
order in eq.~(4.3).  This approximation
corresponds to the simplest physical picture where
the kaon is annihilated by the pseudoscalar quark
current while the pion pair is born by the gluon
operator.  We should not that the above separation
of the operator (4.2) into the quark and gluon
parts leads to some uncertainties in the final
result since the effects of the interaction
between them are lost. However the corresponding
corrections seem to be suppressed at least at the
perturbative level.  Eqs.~(4.1, 4.3-4.6) result
in  the effective chiral Lagrangian of the  form
$$
L_G^{ch}=-{G_F\over\sqrt 2}V_{ud}V^*_{us}(1-\tau)
\left(-{1\over 480\beta_3}{f^4_\pi\over m^2_c}
(U^\dagger \chi)_{23}
tr_{fl}(\partial_\mu U^\dagger\partial^\mu U)\right)
+h.c.
\eqno(4.7)
$$
Since the pion pair is born by the gluon operator
this Lagrangian describes the  investigated decay
channel with gluons forming an intermediate state.

Now the corresponding $K\rightarrow\pi\pi$  decay
amplitude becomes explicitly calculable. We will
use the standard parametrization of the amplitude
$A_0$ with the isospin transfer  $\Delta I=1/2$
$$
ReA_0=
{G_F\over\sqrt 2}\sin\theta_c
\cos\theta_cg_{1/2}f_K m^2_K
\eqno(4.8)
$$
where $\theta_c$ stands for Cabibbo angle,
$g_{1/2}$ is a dimensionless parameter.  The new
contribution reads
$$
\Delta g_{1/2}={1\over 30\beta_3}{m^2_K\over m^2_c}
\sim 10^{-3}.
\eqno(4.9)
$$
While the experiment gives \cite{14}
$$
g^{exp}_{1/2}=3.9
\eqno(4.10)
$$
and the most recent theoretical estimation is
$g_{1/2}\sim 2.4$ \cite{2}.  As we see the local
(perturbative) part of the new decay mode is
negligible according to the general estimate of
the scale of the leading order charmed quark mass
corrections done in Sect.~2.

However the local  effective Hamiltonian (4.1)
does not exhaust the whole physics of the
meson-gluon transitions. It cannot account for
the long-distance contribution connected with the
propagation of the soft $u$-quark  round the loop
of the annihilation diagram. Because of the
lightness of the $u$-quark this contribution can
not be represented as a local vertex. It
ultimately depends on the infrared properties of
{\it QCD} and requires non-perturbative approach.
We will follow the general line of the approach
applied in Sect.~3 to the calculation  of the
penguin operator matrix element.

In so doing we start with a tree level Hamiltonian
which after decoupling of the $c$-quark has the
form
$$
H^{tr}_{\Delta
S=1}= {G_F\over\sqrt 2}V_{ud}V^*_{us}Q_2
+h.c.
\eqno(4.11)
$$
The quantity of interest is an effective theory
realization in terms of mesonic variables of the
part of the operator $Q_2\equiv Q^u_2$ which is
responsible for the kaon transfer into gluons.
Invoking the results of our previous consideration
we can write down this part in the  form
$$
Q^G_2=g^Gf^2_\pi (U^\dagger \chi)_{23}
tr_{fl}(\partial_\mu U^\dagger\partial^\mu U)
\eqno(4.12)
$$
where $g^G$ is a dimensionless parameter to be
computed.  We should note that the chiral
representation of the whole operator $Q_2$
contains a large number of structures but we are
interested only in the part corresponding to the
transition with the gluons forming an intermediate
state which has the unique representation (4.12).
Indeed, there is the single $SU_V(3)$ octet terms
(4.12) in the chiral weak Lagrangian describing
the $K\rightarrow\pi\pi$ decays which is
proportional to the $tr_{fl}(\partial_\mu
U^\dagger\partial^\mu U)$ as it required by
eq.~(4.5) \cite{3}.  Thus the problem is reduced
to the computation  of the chiral coupling
constant $g^G$. As for the parameters $g$ and $g'$
of the chiral representation of the penguin
operator it can be done by studying the
appropriate GF via {\it QCD} sum rules technique.
For the technical reason working with a two point
GF is preferable. In the given decay channel the
pions are born by a gluon cloud therefore the
gluon operator $G^a_{\mu\nu}G^a_{\mu\nu}$ can play
the role of an interpolating operator of the pion
pair. Thus, it is natural to choose GF in the
form
$$
G(p)=\int
\langle 0|T{\alpha_s\over\pi}G^a_{\mu\nu}G^a_{\mu\nu}(x)
Q_2(0)|K^0(q)\rangle e^{ipx}dx|_{q=0}.
\eqno(4.13)
$$

A remark about the chiral limit for the kaon in
eq.~(4.13) is necessary. The representation (4.12)
fixes the correct $O(p^4)$ chiral behavior of the
considered decay amplitude and does not depend
explicitly on the kaon momentum.  Keeping a
non-vanishing kaon momentum leads to a shift of
the decay amplitude that lies beyond the accuracy
of the present approach.  Thus we can put $q=0$ in
eq.~(4.13) and work with GF depending on one
argument only.

Saturating GF (4.13) by the $\pi^+\pi^-$ and
$\pi^0\pi^0$ states (the lowest states with the
proper quantum numbers), substituting the operator
$Q_2$ for its mesonic realization (4.12) and using
the low-energy theorem (4.5) one obtains at small
momentum $p$ the following physical representation
$$
G(p)=g^G{32\over \pi^2\beta_3}{m^2_K\over f\pi}p^4
\ln\left({-p^2\over\mu^2}\right)+O(p^6).
\eqno(4.14)
$$

The theoretical side reads after making use of
{\it OPE}
$$
G(p)=i{1\over 2\pi^2}
\ln\left({-p^2\over\mu^2}\right)
{\alpha_s\over\pi}
\langle 0|m_s\bar s_Rg_sG^a_{\mu\nu}
t^a\sigma_{\mu\nu}d_L|K^0(q)\rangle |_{q=0}
+O(\alpha^2p^2)+O(p^{-2}).
\eqno(4.15)
$$
Factor $m_s$ in eq.~$(4.15)$ provides the correct
chiral property of GF and justified the
representation (4.12) for the operator $Q_2$.  By
contraction of the kaon state  one can transform
eq.~(4.15) into the  expression
$$
G(p)={1\over 4\pi^2}
\ln\left({-p^2\over\mu^2}\right)
{\alpha_s\over\pi}
f_Km^2_Km^2_0 .
\eqno(4.16)
$$

For extracting information about the chiral
coupling constant $g^G$ we use finite energy sum
rules with the result
$$
g^G={3\beta_3\over 128}{f_K\over
f_\pi}{f^2_\pi m^2_0\over s^2_0}{\alpha_s(s_0)\over
\pi}.
\eqno(4.17)
$$

To take into account the strong interaction at
short distances the operator $Q^G_2$ in the
effective Hamiltonian has to be multiplied by the
corresponding Wilson coefficient $z_2(s_0)$.
Finally the new contribution to the theoretical
estimate of the $\Delta I=1/2$ amplitude in terms
of parameter $g_{1/2}$ takes the form
$$
\Delta g_{1/2}=z_2(s_0){3\beta_3\over 8}
{m^2_0m^2_K\over s^2_0}{\alpha_s(s_0)\over \pi}.
\eqno(4.18)
$$
This result needs some comments.

\noindent
1. This next-to-leading in $1/N_c$ expansion
contribution is missed within the factorization
framework and also within any approach where quark
currents in four-quark operators are replaced by
their mesonic counterparts separately.

\noindent
2. The gluon cloud in the intermediate state does
not form a resonance state and, therefore, the
contribution $(4.18)$ is not suppressed by a large
scalar meson mass.

\noindent
3. In general the more complicated scalar
colorless gluon configuration, for example
$f^{abc}G^a_{\mu\nu}G^b_{\nu\lambda}
G^c_{\lambda\mu}$, could be an intermediate state
in this channel as well. However the theorem (4.5)
shows that the two pion form factor of these
configurations can be of $O(p^4)$ or higher order
in chiral expansion and that lead only to the
negligible $O(p^6)$ shift of the decay amplitude.
Indeed, the theorem (4.5) sets the equivalence
between the trace of the energy momentum tensor
expressed in terms of {\it QCD} and mesonic
degrees of freedom.  In chiral limit in the
leading order in momentum expansion there is a
single Lorentz invariant flavor singlet mesonic
configuration which is proportional to the trace
of the energy momentum tensor.  On the other hand
there are no extra gluon contributions to the
conform anomaly.

The question now is what numerical value for the
duality interval $s_0$ has to be used. Actually,
the allowed value of the duality threshold is
quite restricted by the form of the physical
spectrum and by the requirement of absence of
uncontrollable $\alpha_s$ corrections.  To
suppress contributions of higher mass states, for
example, a scalar meson $\sigma (0.9~GeV)$, to the
considered channel one has to take
$s_0<(0.9~GeV)^2$.  At the same time the physical
representation (4.14) is obtained in the leading
order in chiral perturbation theory and the whole
procedure is justified until the ratio
$s_0/\Lambda^2_\chi$ remains small. On the other
hand at the scale $\mu$ less than $0.8~GeV$ the
perturbative $\alpha_s$ corrections to Wilson
coefficients become uncontrollable \cite{2} and
for consistency of the approach one has to set the
low limit of the duality interval to be
$s_0>(0.8~GeV)^2$. The reasonable choice for the
duality interval now reads $s_0=(0.8~GeV)^2$.

Following the approach of the Sect.~3 let us
estimate the uncertainty of our result.  On the
physical side of sum rules the errors related to
higher order terms in chiral expansion, which have
been omitted in eq.~(4.12), are, in general,
unknown. But one can hope that in the spirit of
chiral perturbation theory they are about $25\%$
\cite{20,27}.  On the theoretical side of the sum
rules the errors come from two sources. The first
one is the perturbative part of {\it OPE} (the
unit operator) that is suppressed by a loop factor
$\alpha_s/4\pi\sim 10^{-3}$ and cannot lead to a
sizable change of our result. Next
non-perturbative corrections due to operators with
higher dimensionality seem to be more important.
They start with the dimension eight operators
which have already been discussed. Numerical
estimates would require knowing the matrix
elements of those operators between the kaon and
the vacuum state which are not available now. But
as a first approximation the relative weight of
these corrections can be represented by the ratio
$$
{\langle
{\alpha_s\over\pi}G^a_{\mu\nu}G^a_{\mu\nu}
\rangle\over\Gamma (5)}
:{s_0m_0^2\over 4\pi^2\Gamma (3)}\sim 0.1
\eqno(4.19)
$$
where $\langle
{\alpha_s\over\pi}G^a_{\mu\nu}G^a_{\mu\nu}
\rangle\sim(330~MeV)^4$ \cite{30}, the Gamma
function factor $1/\Gamma (n)$ comes from the
quark loop with the $n-1$ gluon field or mass
insertion. Here we find the same situation as in
the analysis of $1/m_c$ corrections where the
contribution of the dimension eight operators is
suppressed rather numerically than parametrically.

Taking into account the uncertainty of
determination of the parameter $m^2_0$ we estimate
the error bound to be about $40\%$. Numerically
one obtains
$$
\Delta g_{1/2}=0.56\pm 0.22
\eqno(4.20)
$$
at the point $s_0=\mu^2 =(0.8~GeV)^2,~\Lambda_{QCD}=
300~MeV,~z_2(s_0)=1.49~$\cite{9}.

Thus, the new contribution provides about $15\%$
of the experimentally observable amplitude
$(4.10)$. At the same time it is comparable with
the leading order result for the decay amplitude
obtained by the naive factorization of the
four-quark operator $Q_2$  when all strong
interaction corrections are neglected
$$
g_{1/2}^{fac}=5/9.
\eqno(4.21)
$$

This implies the strong violation of factorization
in the $O(p^4)$ order in chiral expansion that
leads to additional enhancement of the theoretical
estimate of the $K\rightarrow\pi\pi$ decay
amplitude with the isospin transfer $\Delta
I=1/2$.

\section{
Conclusion.
}

In the present paper we make an attempt to clarify
the role of annihilation or "penguin" mode in the
descriptions of the $K\rightarrow\pi\pi$ decay in
the Standard Model.  We concentrate our attention
on  some new aspects which have not been
considered yet.

The complete form of the effective low-energy
$\Delta S=1$ Hamiltonian up to the first order in
$m_s$, $1/m_c$ and $\alpha_s$ with $m_t\sim M_W$
is obtained.  The violation of factorization for
the mesonic matrix element of the operator $Q_2$
caused by $K\rightarrow\pi\pi$ transition with the
gluons in the intermediate state is also analyzed.

We propose the regular method to evaluate the
mesonic matrix elements of the effective
Hamiltonian for $K\rightarrow\pi\pi$ transitions.
The weak Hamiltonian is represented through the
variables of chiral effective theory  and the
coefficients of proportionality between
quark-gluon operators and mesonic operators are
determined via {\it QCD} sum rules for an
appropriate three-point GF.  We apply this
technique for computing the {\it QCD} penguin
operator matrix element.

The main results are

\noindent
1. The contribution of the dimension eight
operators (non-leading $1/m_c$ corrections) proves
to be small numerically (about 10 \% of the
leading term) when one  estimates the
corresponding mesonic matrix element within   the
simplest factorization framework.  However its
real magnitude can be estimated only when more
detailed information on the matrix elements  of
the  local operators with  dimension eight
becomes available, for example, from lattice
calculations.

\noindent
2. The quark-gluon operator   $m_s{\bar
s}_Rg_sG_{\mu\nu}\sigma^{\mu\nu}d_L$ contributes
to the imaginary part of the effective Hamiltonian
in the case of a heavy top quark already in the
first order in $\alpha_s$ due to incomplete GIM
cancelation.  However its contribution to the
parameter $\epsilon'$ is  suppressed
because of the smallness of the corresponding
Wilson coefficient and the specific (tadpole)
chiral structure of this operator.

\noindent
3. Sum rules predict for the  matrix element of
the {\it QCD} penguin operator
$$
{1\over i}\langle
\pi\pi|Q_6(1~GeV)|K^0\rangle
=-(0.34\pm 0.09)~GeV^3
$$
where the error bars estimate contributions of
higher orders of the chiral expansion.  The
analysis carried out above shows that the
uncertainties coming from other sources are
relatively small.  This value is somewhat larger
than the results obtained within the $1/N_c$
approximation framework \cite{4,5} but it is
somewhat smaller than optimistic estimate of ref.
\cite{7}.  Our result gets also into the interval
given by lattice models \cite{19}.  The relatively
large value of the matrix element along with the
strong enhancement of the corresponding Wilson
coefficient \cite{2} make the penguin operator
quite important in the analysis of the $\Delta
I=1/2$ rule since it provides about $20\%$ of the
decay amplitude with $\Delta I=1/2$.

\noindent
4.
The non-perturbative contribution to the  mesonic
matrix element of the leading operator $Q_2$
produced by the transitions with gluons in the
intermediate state gives an additional enhancement
of the theoretically predicted
$K\rightarrow\pi\pi$ decay amplitude with $\Delta
I=1/2$ and provides about $15\%$ of the
experimentally observable amplitude value.  New
contribution is of $O(p^4)$ and is lost within the
factorization framework but numerically is
comparable with the result of  naive factorization
for the $\Delta I=1/2$ decay amplitude.  It allows
us to conclude that there is a sizable violation
of the factorization in the $O(p^4)$ order in
chiral expansion.

To conclude, the comparison between the
"penguin-like" and "non-penguin" parts of the
$\Delta I=1/2$ amplitude is quite instructive.
According to the recent work \cite{2} the
current-current operators $Q_1$ and $Q_2$ (without
$Q^G_2$ part) give about $45\%$ of the
experimentally observable amplitude while the
total effect of the "penguin-like" contributions
of the operators $Q_6$ and $Q^G_2$ turns out to be
about $30-40\%$.  So large "penguin-like"
contribution improves considerably the theoretical
description of the $\Delta I=1/2$ rule in
non-leptonic kaon decays.  Indeed, our results
imply $ReA^{th}_0\sim 0.8ReA^{exp}_0$ while ref.
\cite{2} gives $ReA^{th}_0\sim 0.6ReA^{exp}_0$
with penguin contribution being only about $15\%$.
As we see, the discrepancy between theory and
experiment still exists but it is getting smaller.
This result can be presented in more impressive
form if one considers the ratio $ReA_0/ReA_2$
where $A_2$ stands for the amplitude with the
isospin transfer  $\Delta I=3/2$ \cite{2}. Then we
obtain $(ReA_0/ReA_2)^{th}\sim 21.$ that is very
close to the experimental data
$(ReA_0/ReA_2)^{exp}=22.3$.  We should emphasize
that this result is obtained without the standard
practice of using the extremely low normalization
point or extremely light strange quark. Since the
resources of the perturbation theory seem to be
exhausted the further progress in the theoretical
explanation of the $\Delta I=1/2$ problem will
probably be connected with studying the
unfactorizable contributions to the mesonic matrix
elements caused by strong interactions at low
energy.

\vspace{1.0cm}
\noindent
{\large \bf Acknowledgments}

\vspace{0.5cm}
\noindent
A.A.Pivovarov is thankful to K.Higashijima,
Y.Okada, M.Tanabashi and A.Ukawa for interesting
discussions and to all colleagues in theory Group
for the great hospitality extended to him during
the stay at KEK.  This work is supported in part
by Russian Fund for Fundamental Research under
Contract No.  93-02-14428, 94-02-06427, by Soros
Foundation and by Japan Society for the Promotion
of Science (JSPS).


\end{document}